\documentclass[24pt]{article}
\usepackage{amssymb}
\usepackage{amsfonts}
\usepackage{amssymb,amsmath}
\usepackage{mathrsfs}
\usepackage{latexsym}
\usepackage{amsmath}
\usepackage{latexsym}
\usepackage[cp1251]{inputenc}
\usepackage{graphicx}
\usepackage{color}

\title{Finite periodic $\delta-\delta'$ comb}
\author{O. I. Hryhorchak, V. S. Pastukhov\\
{\small Department for Theoretical Physics, Ivan Franko National
University of Lviv,}\\
{\small 12, Drahomanov Str., Lviv, UA--79005,
Ukraine}\\
\small{\it{Orest.Hryhorchak@lnu.edu.ua}}}

\def\ch{\mathop{\rm ch}\nolimits}
\def\sh{\mathop{\rm sh}\nolimits}

\def\ctg{\mathop{\rm ctg}\nolimits}

\begin{document}
\renewcommand{\abstractname}{Abstract}
\maketitle

\begin{abstract}
A finit periodic $\delta-\delta'$ comb was solved by the help of both classical approach based on a direct solving of a Sr\"{odinger} equation and a quantum wave impedance method. It was demonstrated that the violation of a periodicity
leads to the formation of energy levels in band gaps.  The expresion for a surface levels (Tamm's levels) were found in this model system. Obtained results allow  extending the scope of theoretical models which are applicable for a describing real quantum mechanical systems as well as better understanding the physical properties of systems in which the boundary dependent states are involved, especially the influence of boundary dependent electronic states on the physical properties of low-dimensional systems.
\end{abstract}

\section{Introduction}
In the pioneer article \cite{Hague_Hague_Khan:1990} authors using a quantum wave impedance concept developed a method for calculation of energy bands for periodic structures and demonstrated that structures with a non-rectangular potential distribution can be analysed within this approach. Thus, they sketched the method for the eigenstates finding in  periodic structures with an arbitrary potential variation within the unit cell. 
Later it was shown how with the help of a quntum wave impedance technique to describe crystal-like structures like electromagnetic, phonon and photon crystals \cite{Nelin:2004, Nelin:2005, Nelin:2006, Nelin:2009_2, Khatyan_Gindikina_Nelin:2015, Nazarko_etall:2009, Gindikina_Zinger_Nelin:2015, Nelin:2007_1, Nelin_Nazarko:2012, Babushkin_Nelin:2011, Nazarko_etall:2011, Nelin:2012, Nazarko_etall:2011_1, Nazarko_Timofeeva_Nelin:2010, Nelin_Zinher_Popsui:2017, Nelin:2004_1, Nazarko_etall:2015}. In that papers a zone diagram formation in periodic systems was described and it was discussed how these results  can be used for a design of nano-electronic devices.

In our previous papers \cite{Arx1:2020, Arx2:2020, Arx3:2020, Arx4:2020, Arx5:2020, Arx6:2020, Arx8:2020, Arx10:2020} we demonstarted how a quantum wave impedance method can be applied to the investigation of infinite and semi-infinite periodic structures which contain both a piesewise constant potential and zero-range singular potentials, namely $\delta$ and $\delta-\delta'$ while in
\cite{Arx7:2020} we developed an approach to a study of systems with a complicated geometry of a potential. Next step we have done in \cite{Arx9:2020}, where a technique of theoretical study of finit periodic structures using the advantures of both transfer matrix approach and a quantum wave impedance method was proposed. 

In this paper we consider the extended finite Dirac comb model, namely, the model given by a finite number $N$ of equally spaced $\delta-\delta'$ functions restricted by edges of a height $U_L$ from the left side and $U_R$ from the right side. One can treat it as the extension of a Sokolov's model \cite{Sokolow:1934}. As it was mentioned in  \cite{Arx4:2020} there is only one preprint \cite{Gadella:2019} dedicated to the infinite $\delta-\delta'$ comb. And there are no one which would describe a finite $\delta-\delta'$ comb.  So that, this model will enrich the scope of models of finite periodic systems.  

But before proceeding it is worth to discuss the introdunction of a $\delta'$-potential. Although this extend the scope of physical systems which can be modelled by  using of this potential (compared to $\delta$-potential only) \cite{Zolotaryuk_Zolotaryuk:2015, Zolotaryuk_Tsironis_Zolotaryuk:2019} but at the same time, since a $\delta'$-potential
causes not only a break of a first derivative of a wave function (like a $\delta$-potential) but also a break of a wave function itself, there are several approaches for its definition \cite{Coutinho_Nogami_Perez:1997, Coutinho_etall:1999, Coutinho_Nogami_Toyama:2009, Patil:1994, Christiansen_etall:2012, Zolotaryuk_Christiansen_Iermakova:2006, Golovaty2012, Golovaty_Hryniv:2012}. Each of these approaches are related to different physical realities \cite{Lange:2015} and bear different values of reflection and transmission coefficients \cite{Christiansen_etall:2003,Toyama:2007,Christiansen_etall:2012, Ahmed_etall:2016}. 

Tis problem was mathematically clarified in \cite{Kurasov:1996, ALBEVERIO_DABROWSKI_KURASOV:1998}. Their approach gives the relation
between a $\delta'$ potential and matching conditions at the origin. Here we will follow this approach and use the matching condition for a quntum wave impedance in a case $\delta-\delta'$ potential which was derived in  \cite{Arx4:2020}. 

\section{Finite $\delta-\delta'$ comb. Quantum wave impedance approach}
The potential energy of this system has a form:
\begin{eqnarray}\label{dd'pot}
U(x)\!=\!U_L\theta(-x)\!-\! \sum_{n=1}^N\alpha\delta(x-nl)\!+\!\beta\delta'(x-nl)
+U_R\theta(x-L),
\end{eqnarray}
where $\theta(x)$ is a Heaviside step function, $l$ is the distance between adjacent $\delta-\delta'$ functions, $L=(N+1)l$ is the width of the studied system, $m$ is the mass of a particle and $E$ is its energy.

Using the matching condition \cite{Arx4:2020}, namely
\begin{eqnarray}
Z(0+)=-\frac{i\hbar\tilde{\alpha}}{m(1+\tilde{\beta})^2}+
\frac{(1-\tilde{\beta})^2}
{(1+\tilde{\beta})^2}Z(0-)
\end{eqnarray}
for an elementary cell of a $\delta-\delta'$ comb we have
\begin{eqnarray} 
Z(a)\!\!\!&=&\!\!\!z_0\frac{(1+\tilde{\beta})^2}
{(1-\tilde{\beta})^2}\frac{Z(b)\ch[i\xi]-z_0\sh[i\xi]}{z_0\ch[i\xi]-Z(b)\sh[i\xi]}+\frac{i\hbar\tilde{\alpha}}{m(1-\tilde{\beta})^2}=\nonumber\\
\!\!\!&=&\!\!\!\frac{z_0}{(1-\tilde{\beta})^2}\left[\frac{}{}
z_0\!\left(\!2i\Omega\ch[i\xi]\!-\!(1\!+\!\tilde{\beta})^2\!\sh[i\xi]\right)\!\!\right.-\nonumber\\
\!\!\!&-&\!\!\!\left.\!Z(a)\!\left(\!2i\Omega\!\sh[i\xi]\!-\!(1\!+\!\tilde{\beta})^2\!\ch[i\xi]\!\right)\right]\!\!\left[\frac{}{}
z_0\ch[i\xi]-Z(a)\sh[i\xi]\right]^{-1}.
\end{eqnarray}
On the base of relations obtained in \cite{Arx9:2020} we get 
\begin{eqnarray}
Z_{11}&=&2i\Omega\ch[i\xi]-(1+\tilde{\beta})^2\sh[i\xi],\nonumber \\
Z_{12}&=&2i\Omega\sh[i\xi]-(1+\tilde{\beta})^2\ch[i\xi],\nonumber \\
Z_{21}&=&(1-\tilde{\beta})^2\ch[i\xi],\nonumber\\
Z_{22}&=&(1-\tilde{\beta})^2\sh[i\xi].
\end{eqnarray}
Before proceeding we have to check if $\det|Z|=1$. So,
\begin{eqnarray}
\det|Z|=Z_{11}Z_{22}-Z_{12}Z_{21}=\left(1-\tilde{\beta}^2\right)^2.
\end{eqnarray}
It means that we have to normalize elements $Z_{ij}$ and after the procedure of normalization we get
\begin{eqnarray}
Z_{11}&=&\frac{2i\Omega}{1-\tilde{\beta}^2}\ch[i\xi]-\frac{1+\tilde{\beta}}{1-\tilde{\beta}}\sh[i\xi],\nonumber \\
Z_{12}&=&\frac{2i\Omega}{1-\tilde{\beta}^2}\sh[i\xi]-\frac{1+\tilde{\beta}}{1-\tilde{\beta}}\ch[i\xi],\nonumber \\
Z_{21}&=&\frac{1-\tilde{\beta}}{1+\tilde{\beta}}\ch[i\xi],\nonumber\\
Z_{22}&=&\frac{1-\tilde{\beta}}{1+\tilde{\beta}}\sh[i\xi].
\end{eqnarray}
Then using the relations between a transfer matrix and a quantum wave impedance matrix \cite{Arx9:2020} we obtain the elements of a transfer matrix for an elementary cell of a $\delta-\delta'$ comb 
\begin{eqnarray}
T_{11}&=&
\frac{1+\tilde{\beta}^2}{1-\tilde{\beta}^2}\left(1+\frac{i\Omega}{1+\tilde{\beta}^2}\right)e^{-i\xi}
,\nonumber\\
T_{12}&=&\frac{i\Omega-2\tilde{\beta}}{1-\tilde{\beta}^2}e^{i\xi},
\nonumber\\
T_{21}&=&
-\frac{i\Omega+2\tilde{\beta}}{1-\tilde{\beta}^2}e^{-i\xi},\nonumber\\
T_{22}&=&\frac{1+\tilde{\beta}^2}{1-\tilde{\beta}^2}\left(1-\frac{i\Omega}{1+\tilde{\beta}^2}\right)e^{i\xi}.
\end{eqnarray}
Thus, the transfer matrix of an elementary cell of $\delta-\delta'$ comb has the following form 
\begin{eqnarray}
T\!=\!\frac{1}{1\!-\!\tilde{\beta}^2}\!\!
\left(\begin{array}{cc}
\left(1+\tilde{\beta}^2+i\Omega\right)e^{-i\xi}& (i\Omega -2\tilde{\beta})e^{i\xi}\\
-(i\Omega+2\tilde{\beta}) e^{-i\xi} &  \left(1+\tilde{\beta}^2-i\Omega\right)e^{i\xi}
\end{array}\right).
\end{eqnarray} 
If we put $\tilde{\beta}=0$ we get a transfer matrix for a Dirac comb. The sum of $T_{11}/2$ and $T_{22}/2$ gives us the dispersion relation for a $\delta-\delta'$ comb. Now our task is to find the energies of surface states for this model. For this we rewrite a formula for a Dirac comb \cite{Arx9:2020} in the following way (assuming for the simplification that $U_L=U_R$)
\begin{eqnarray}
&&\sin(\chi)\ctg((N+1)\chi)-\cos(\chi)=(r_0-r_N)(1-\tilde{\beta}^2)\times\nonumber\\
&&\times\left(r_N(T_{11}^0+\tilde{\beta}^2e^{-i\xi})+r_0r_N(T_{12}^0-2\tilde{\beta} e^{i\xi})\right.-\\
&&\left.-(T_{21}^0-2\tilde{\beta} e^{-i\xi})-r_0(T_{22}^0+\tilde{\beta}^2e^{i\xi})\right)^{-1},
\end{eqnarray}
where index $0$ on the top of elements $T_{ij}$ means the matrix elements of the transfer matrix for an elementary cell of a Dirac comb. Now taking into account that
\begin{eqnarray}
r_N\tilde{\beta}^2e^{-i\xi}+2r_0r_N\tilde{\beta} e^{i\xi}-2\tilde{\beta} e^{-i\xi}-r_0\tilde{\beta}^2e^{i\xi}=\frac{4\tilde{\beta}^2z_0z_ke^{-i\xi}}{z_0^2-z_k^2}
\end{eqnarray}
we obtain the expression for a determination of energies of surface states in a $\delta-\delta'$ finite comb 
\begin{eqnarray}
\sin(\chi)\ctg((N+1)\chi)-\cos(\chi)=
\frac{1-\tilde{\beta}^2}{2}\frac{(\varkappa_E^2L^2/\xi^2-1)\sin(\xi)+2\varkappa_EL/\xi\cos(\xi)}{\Omega-(1-\tilde{\beta}^2)\varkappa_EL/\xi}.
\end{eqnarray} 

\section{Finit $\delta-\delta'$ comb. Classical approach}
After an application of a quantum wave impedance approach it will be instructive and useful to solve the model of $\delta-\delta'$ comb using a Sr\"{o}dinger equation directly. Its general solution with a potential (\ref{dd'pot}) is well-known besides $x=nl$ points, where $n=1,\ldots,N$. So, in any n-th cell we have the following wave function
\begin{eqnarray}\label{psinAB}
\!\!\!\psi_n(x)\!=\!A_n\sin[k_0(x\!-\!nl)]\!+\!B_n\cos[k_0(x\!-\!nl)], nl<x<(n\!+\!1)l,
\end{eqnarray}
where $k_0=\sqrt{2mE}/\hbar$.
Outside the region of $\delta-\delta'$-comb we have wave functions in the following form:
\begin{eqnarray}
\psi_L(x)&=&D_Le^{\varkappa_L x}, \qquad x\leq 0,\nonumber\\
\psi_R(x)&=&D_Re^{-\varkappa_R (x-L)}, \qquad x\geq L,\nonumber
\end{eqnarray}
where $ \varkappa_L=\sqrt{2m(U_L-E)}/\hbar$,$ \varkappa_R=\sqrt{2m(U_R-E)}/\hbar$.

Now it is time to consider wave-functions $\psi_n(x)$ at the vicinity of points $x=nl$. We deal with such wave-functions which belong to the subspace of
a self-adjoint extension of a kinetic energy operator: $-\frac{\hbar^2}{2m}\frac{d^2}{dx^2}$.
These functions, as it was well-established earlier \cite{Gadella:2009}, have to satisfy matching conditions at the origine of a $\delta-\delta'$ potential:
\begin{eqnarray}
\begin{pmatrix}
\psi_{n+1}((n+1)l+0) 
\\
\psi_{n+1}'((n+1)l+0) 
\end{pmatrix}
=
\begin{pmatrix}
\frac{1+\tilde{\beta}}{1-\tilde{\beta}}, \qquad 0 
\\
-\frac{\tilde{\alpha}}{1-\tilde{\beta}^2}, 
\frac{1-\tilde{\beta}}{1+\tilde{\beta}} 
\end{pmatrix}
\begin{pmatrix}
\psi_{n}((n+1)l-0) 
\\
\psi_{n}'((n+1)l-0) 
\end{pmatrix},
\end{eqnarray}
where
\begin{eqnarray}
\tilde{\beta}=\frac{m{\beta}}{\hbar^2},\qquad \tilde{\alpha}=\frac{2m{\alpha}}{\hbar^2}.
\end{eqnarray}
So, if we apply written above conditions to the expression (\ref{psinAB}) it will give us the system of equations for
unknown coefficients $A_n$ and $B_n$ in the following form:
\begin{eqnarray}
B_{n+1}&=&\frac{1+\tilde{\beta}}{1-\tilde{\beta}}\left[\frac{}{}A_n\sin(\xi)+B_n\cos(\xi)\right],\nonumber\\
A_{n+1}&=&\frac{1-\tilde{\beta}}{1+\tilde{\beta}}\left[\frac{}{}A_n\cos(\xi)-B_n\sin(\xi)\right]-\frac{\tilde{\alpha} l B_{n+1}}{\xi(1+\tilde{\beta})^2},
\end{eqnarray}
where $\xi=k_0l$. 
We obtained the difference equation. With a common assumption that the solutions for $A_n$ and $B_n$ are of the form 
$A_n=C_Ae^{i\lambda n}$, $B_n=C_Be^{i\lambda n}$ for $C_A$ and $C_B$ we get
\begin{eqnarray}\label{seCaCb}
\!\!\!\!\!\!C_Be^{\pm\lambda}&=&\frac{1+\tilde{\beta}}{1-\tilde{\beta}}\left[\frac{}{}C_A\sin(\xi)+C_B\cos(\xi)\right],\nonumber\\
\!\!\!\!\!\!C_Ae^{\pm\lambda}&=&\frac{1-\tilde{\beta}}{1+\tilde{\beta}}\left[\frac{}{}C_A\cos(\xi)-C_B\sin(\xi)\right]-
\frac{\tilde{\alpha} l C_Be^{\pm\lambda}}{\xi (1+\tilde{\beta})^2}
\end{eqnarray}
or
\begin{eqnarray}
\!\!\!\!\!\!&&\!\!\!\!\!\!\frac{1+\tilde{\beta}}{1-\tilde{\beta}} C_A\sin(\xi)+C_B\left(\frac{1+\tilde{\beta}}{1-\tilde{\beta}}\cos(\xi)-e^{\pm i\lambda}\right)=0,\nonumber\\
\!\!\!\!\!\!&&\!\!\!\!\!\!C_A\!\left(\frac{1-\tilde{\beta}}{1+\tilde{\beta}}\cos(\xi)-e^{\pm i\lambda}\right)\!-\!\frac{1-\tilde{\beta}}{1+\tilde{\beta}} C_B\!\left(\sin(\xi)
\!+\!\frac{\tilde{\alpha} l}{\xi(1-\tilde{\beta}^2)}e^{\pm i\lambda}\right)\!=\!0,\nonumber\\
\end{eqnarray}
A non-trivial solution of this system demands its determinant to be equal zero.  
Consequently, we obtain the equation for $\lambda$:
\begin{eqnarray}
\left(\frac{1+\tilde{\beta}}{1-\tilde{\beta}}\cos(\xi)-e^{\pm i\lambda}\right)
\left(\frac{1-\tilde{\beta}}{1+\tilde{\beta}}\cos(\xi)-e^{\pm i\lambda}\right)+\sin(\xi)\left(\sin(\xi)
+\frac{\tilde{\alpha} l}{\xi(1-\tilde{\beta}^2)}e^{\pm i\lambda}\right)=0
\end{eqnarray}
or after simple transformations we get an equation:
\begin{eqnarray}
e^{\pm2i\lambda}+1-2e^{\pm i\lambda}\left(\frac{1+\tilde{\beta}^2}{1-\tilde{\beta}^2}\cos(\xi)-
\frac{\tilde{\alpha} l}{2\xi(1-\tilde{\beta}^2)}\sin(\xi)\right)=0.
\end{eqnarray}
Taking into account that
\begin{eqnarray}
e^{\pm2i\lambda}+1=2e^{\pm\lambda}\cos(\lambda)
\end{eqnarray}
we finally obtain: 
\begin{eqnarray}\label{dr_dd'}
\cos\lambda=\frac{1+\tilde{\beta}^2}{1-\tilde{\beta}^2}\left(\cos(\xi)-\frac{p}{(1+\tilde{\beta}^2)\xi}\sin(\xi)\right),
\end{eqnarray}
where
$p=\tilde{\alpha} l/2=m\alpha l/\hbar^2$.

Of course, we can represent $\lambda$ as a product of a modulus of a quasi wave-vector $k$ and a width of an elementary cell $l$, namely $\lambda=kl$. This result is the same one obtained in the paper \cite{Gadella:2019}.
In the limit of $\tilde{\beta}=0$ we have a well-known dispersion relation for a Dirac comb, which we derived earlier.

\section{Wave functions of surface states of a finite $\delta-\delta'$ comb}
In this section our task is to derive wave functions of surface states for a finite $\delta-\delta'$ comb. For this we return to the system of equations (\ref{seCaCb}) and find that
\begin{eqnarray}
C_A=\left(\frac{1-\tilde{\beta}}{1+\tilde{\beta}} \frac{e^{\pm\lambda}}{\sin(\xi)}-\ctg(\xi)\right)C_B.
\end{eqnarray}
Using the previous relation and (\ref{psinAB}) we get
\begin{eqnarray}
\psi(x)&=&C_Be^{\pm i\lambda n}\left[\left(\frac{1-\tilde{\beta}}{1+\tilde{\beta}} \frac{e^{\pm\lambda}}{\sin(\xi)}-\ctg(\xi)\right)\sin[k_0(x-nl)]+\right.\nonumber\\
&+&\left.\cos[k_0(x-nl)]\frac{}{}\right]=
\frac{C_Be^{\pm i\lambda n}}{\sin(\xi)}\left\{e^{\pm i\lambda}\frac{1-\tilde{\beta}}{1+\tilde{\beta}}\sin[k_0(x-nl)]+\sin[k_0((n+1)l-x)]\frac{}{}\right\}.
\end{eqnarray}
Notice that $\lambda$ is defined by an equation (\ref{dr_dd'}) and a general solution is the superposition of solutions with $+\lambda$ and $-\lambda$:
\begin{eqnarray}
\psi(x)\!\!\!\!
&=&\!\!\!\!\frac{C_B^+e^{i\lambda n}}{\sin(\xi)}\left\{e^{ i\lambda}\frac{1-\tilde{\beta}}{1+\tilde{\beta}}\sin[k_0(x-nl)+\sin[k_0((n+1)l-x)]\right\}+\nonumber\\
\!\!\!\!&+&\!\!\!\!\frac{C_B^-e^{-i\lambda n}}{\sin(\xi)}\left\{e^{ -i\lambda}\frac{1-\tilde{\beta}}{1+\tilde{\beta}}\sin[k_0(x-nl)+\sin[k_0((n+1)l-x)]\right\},
\end{eqnarray}
where $C_B^+$ and $C_B^-$ are two arbitrary constants to be determined by  boundary conditions. Introducing constants $D_1$ and $D_2$
\begin{eqnarray}
D_1=\frac{C_B^++C_B^-}{\sin(\xi)}\qquad D_2=\frac{i(C_B^+-C_B^-)}{\sin(\xi)} 
\end{eqnarray}
we get
\begin{eqnarray}
\psi_n(x)\!\!\!\!&=&\!\!\!\!\frac{1-\tilde{\beta}}{1+\tilde{\beta}}\left(\frac{}{}D_1\cos[(n+1)\lambda]+D_2\sin[(n+1)\lambda]\right)\sin[k_0(x-nl)]+\nonumber\\
\!\!\!\!&-&\!\!\!\!\left(D_1\cos[n\lambda]+D_2\sin[n\lambda]\frac{}{}\right)\sin[k_0(x-(n+1)l)].
\end{eqnarray}
In the next section we will see that assuming $U_L=U_R$ gives 
\begin{eqnarray}
D_1=\frac{D_L}{\sin(\xi)},\qquad
D_2=\frac{D_L\eta(\gamma)}{\sin(\xi)}
\end{eqnarray}
and
\begin{eqnarray}
D_R=\frac{D_L}{\gamma}\left(\frac{}{}\cos[(n+1)\lambda]+\eta(\gamma)\sin[(n+1)\lambda]\right),
\end{eqnarray}
where $\gamma$ and $\eta(\gamma)$ (\ref{eta_gamma}) will be introduced in the next section. Thus, the wave functions of surface states have the following form
\begin{eqnarray}
\!\!\!\!\!&&\Psi_n(x)\!=\!D_L\!\left\{\exp[\varkappa x]\theta(-x)\!+\!
\left(\frac{\cos[(n+1)\lambda]\!+\!\eta(\gamma)
	\sin[(n+1)\lambda]}{\gamma\sin(kl)}\times\right.\right.\nonumber\\
\!\!\!\!\!&&\left.\times\sin[k(x-nl)]-\frac{\cos[n\lambda]+\eta(\gamma)\sin[n\lambda]}{\sin(kl)}\sin[k(x-(n+1)l)]\right)\times
\nonumber\\
\!\!\!\!\!&&\times
(\theta(x-nl)-\theta(x-(n+1)l))+
\frac{1}{\gamma}\left(\frac{}{}\cos[(n+1)\lambda]+\right.
\nonumber\\
\!\!\!\!\!&&\left.\left.+\eta(\gamma)\sin[(n+1)\lambda]\frac{}{}\right)\exp[\varkappa (x-L)]\theta(x-L)
\frac{}{}\right\},
\end{eqnarray}
where $D_L$ can be found from the condition
\mbox{$\int\limits_0^\infty |\Psi_n(x)|^2 dx =1.
	$}

\section{Surface states in a finite $\delta-\delta'$ comb. Classical approach}
In previous sections using a quantum wave impedance approach we have found the relation which allows determining the energies of surface states. In this section we are going to fulfil the same task but with the help of a classical approach.

Let's consider the wave functions of a $\delta-\delta'$ comb at the interfaces of the left-side and right-side edges ($x=0$ and $x=L$). The matching conditions for a wave function $\psi_0(x)$ at a point $x=0$ and for a wave-function $\psi_{n+1}(x)$ at a point $x=L$ are common.
Outside the $\delta$-$\delta'$ comb we have the wave functions $\psi_L(x)\:(x<0)$ and $\psi_R(x)\:(x>L)$, which we have found in the previous section.

So after introducing a notation $\gamma=(1+\tilde{\beta})/(1-\tilde{\beta})$ we get the following relations at a point $x=0$ (n=0):
\begin{eqnarray}
D_1\sin(\xi)=D_L,\quad
D_1\cos(\lambda)+D_2\sin(\lambda)-\gamma D_1\cos(\xi)=\frac{\gamma L\varkappa_L D_L}{\xi}.
\end{eqnarray}
At a point $x=L$ (n=N) we have
\begin{eqnarray} 
&&\left(\frac{}{}D_1\cos[(N+1)\lambda]+D_2\sin[(N+1)\lambda]\right)\sin(\xi)=\gamma D_R,\nonumber\\
&&\left(\frac{}{}D_1\cos[(N+1)\lambda]+D_2\sin[(N+1)\lambda]\right)\cos(\xi)-\nonumber\\
&&-\gamma\left(\frac{}{}D_1\cos(N\lambda)+D_2\sin(N\lambda)\right)=-\frac{\gamma\varkappa_R D_R}{k},
\end{eqnarray}
which for $D_1$ and $D_2$ gives the following relations
\begin{eqnarray}\label{sqD1D2}
&&D_1\left(\gamma\cos(\xi)+\frac{\gamma l\varkappa}{\xi}\sin(\xi)-\cos(\lambda)\right)=D_2\sin(\lambda),\nonumber\\
&&D_1\left(\left(\cos(\xi)+\frac{l\varkappa }{\xi}\sin(\xi)\right)\cos[(N+1)\lambda]-\gamma\cos(Nl)\right)=\nonumber\\
&&D_2\left(\left(\cos(\xi)+\frac{l\varkappa }{\xi}\sin(\xi)\right)\sin[(N+1)\lambda]-\gamma\sin(Nl)\right).
\end{eqnarray}
For the simplification we assume that
\begin{eqnarray}
\varkappa_L=\varkappa_R=\varkappa.
\end{eqnarray}
Taking into account that
\begin{eqnarray}
\cos(n\lambda)=\cos[(n+1)\lambda]\cos(\lambda)+\sin[(n+1)\lambda]\sin(\lambda),\nonumber\\
\sin(n\lambda)=\sin[(n+1)\lambda]\cos(\lambda)-\cos[(n+1)\lambda]\sin(\lambda)
\end{eqnarray}
we obtain:
\begin{eqnarray}
&&D_1\left(\frac{}{}\gamma\cos(\xi)+\frac{\gamma l\varkappa}{\xi}\sin(\xi)-\cos(\lambda)\right)=D_2\sin(\lambda),\nonumber\\
&&D_1\left(\left(\cos(xi)+\frac{l\varkappa}{\xi}\sin(\xi)-\gamma\cos(\lambda)\right)\ctg[(n+1)\lambda]
-\gamma\sin(\lambda)\right)=\nonumber\\
&&D_2\left(\cos(\xi)+\frac{l\varkappa}{\xi}\sin(\xi)-\gamma\cos(\lambda)-\gamma\ctg[(n+1)\lambda]\sin(\lambda)\right)
\end{eqnarray}
or
\begin{eqnarray}
D_1=D_2\eta(\gamma),\quad
D_1\left(\eta(\gamma^{-1})-\ctg[(n+1)\lambda]\right)=
D_2\left(1+\ctg[(n+1)\lambda]\eta(\gamma^{-1})\right),
\end{eqnarray}
where
\begin{eqnarray}\label{eta_gamma}
\eta(\gamma)=\frac{\sin(\lambda)}{\gamma\cos(\xi)+\frac{\gamma l\varkappa}{\xi}\sin(\xi)-\cos(\lambda)},\quad
\eta(\gamma^{-1})=\frac{\gamma\sin(\lambda)}{\cos(\xi)+\frac{l \varkappa}{\xi}\sin(\xi)-\gamma\cos(\lambda)}.
\end{eqnarray}
For a non-trivial solution of (\ref{sqD1D2}) the following relation has to be met
\begin{eqnarray}
\ctg[(n+1)\lambda]=\frac{\eta\left(\gamma\right)\eta\left(\gamma^{-1}\right)-1}{\eta\left(\gamma\right)+\eta\left(\gamma^{-1}\right)}.
\end{eqnarray}
It is the relation for a determination of energies of surface states in the finite $\delta-\delta'$ comb.
If we put $\tilde{\beta}=0$ in this expression then $\gamma=1$, and we get the same result as for a finite Dirac comb  \cite{Arx9:2020}.

The results of numerical calculations for surface states in the first and the second borbidden zones are represented on Figure 5.1. The dependance of a value $\xi=k_0L$ (where $k_0=\sqrt{2E/m_0}$, $m_0$ is a mass of an electron) on parameter $\tilde{\beta}$ at $p=1$ is depicted on Figure 1.
\begin{figure}[h!]
	\centerline{\includegraphics[clip,scale=1.05]{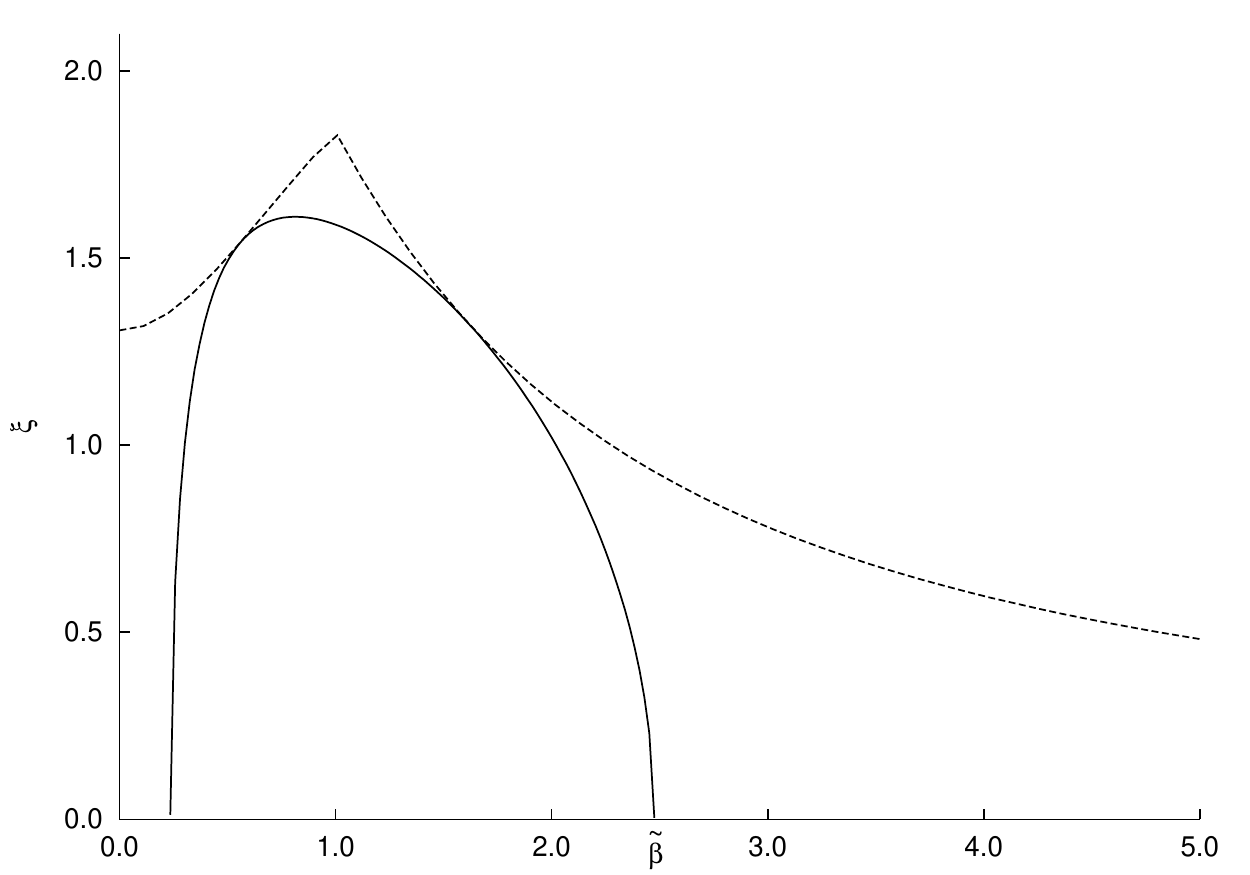}}
	\vspace{-0.3cm}
	\centerline{\includegraphics[clip,scale=1.05]{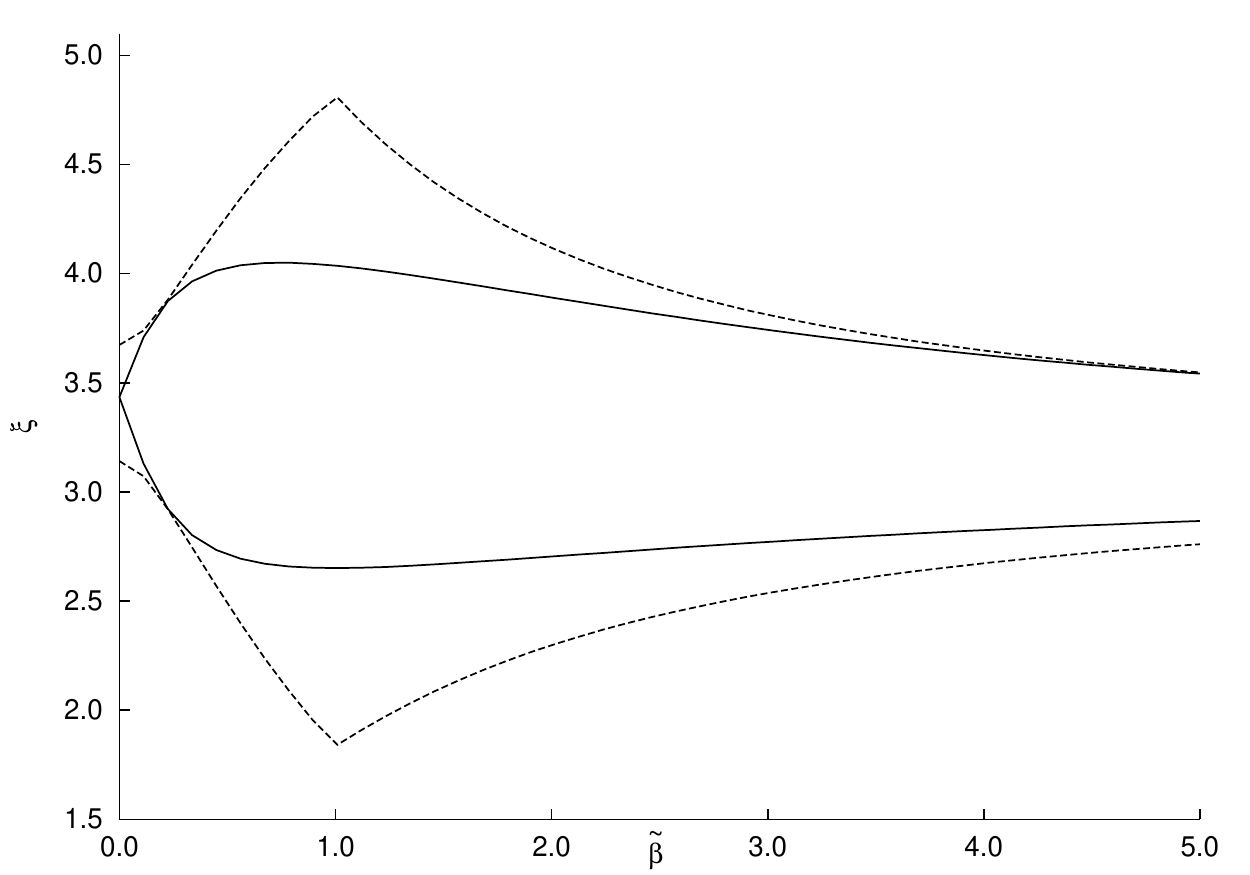}}
	\vspace{-0.1cm}
	\caption{\small{Dependance of a value $\xi=k_0L$ for the surface states in the first (the upper Figure) and in the second (the bottom Figure) forbidden zone on parametr $\tilde{\beta}$ at $p=1$ (solid line). Dashed lines is for the bottom and upper boundaries of the first and the second forbidden zone.}}
\end{figure}

 \newpage
\section{Conclusions}
A $\delta$ potential is
widely discussed in a literature \cite{Demkov_Ostrovskii:1988, Albeverio_etall:2005} as a tool for modelling of real quantum mechanical systems (see, for example \cite{Sfiat:2013}) which allows getting exact solutions. Ofcourse, an inrdoducing a $\delta'$ potential gives new possibilities to ``cath'' main peculiarities of real systems through an investigation of theoretical models.

In this paper we have solved a finit periodic $\delta-\delta'$ comb using both classical approach and a quantum wave impedance method which we developed in  \cite{Arx9:2020}. We both demonstrated that the violation of a periodicity
leads to the formation of energy levels in band gaps and found the expresion for a surface levels (Tamm's levels) in this model system. Our consideration allows to notice that as the number of identical regions increases, the propagating waves give rise to pass bands, while the
standing waves give rise to gaps in the energy spectrum of the system. So we get an energy zone structure (an alternation of bands and gaps) in a finite periodic medium.

Concluding it worth to say that such models help one to understand the physical properties of systems in which the boundary dependent states are involved. First of all the mechanism of how the existence of boundary dependent electronic states influences on the physical properties of low-dimensional systems.
But actually beyond the simplified models, there is much that we do not understand. Thus, finding methods for studying more complicated models of finite periodic systems is very important task. And an impedance concept is a very natural approach in this case since impedance as a response of a medium to a
wave propagation characterizes the wave properties of
studied systems in a generalized manner and allows  analysing the formation of spectral properties and how the parameters and defects which are present in a system effect these properties.

Besides $\delta$ and $\delta'$ potentials one can face also with a $\delta''$-potential in the literature \cite{Patil:1994, Zolotaryuk:2015}. So it would be interesting to consider $\delta-\delta'-\delta''$ comb which we plan to perform in our next paper.

\renewcommand\baselinestretch{1.0}\selectfont


\def\name{\vspace*{-0cm}\LARGE 
	Bibliography\thispagestyle{empty}}
\addcontentsline{toc}{chapter}{Bibliography}

{\small

	\bibliographystyle{gost780u}
	\bibliography{full.bib}
	
}

\newpage

\end{document}